\pdfoutput=1
\documentclass[conference]{IEEEtran}
\IEEEoverridecommandlockouts
\usepackage{cite}
\usepackage{amsmath,amssymb,amsfonts}

\usepackage{graphicx}
\usepackage{algorithm}
\usepackage{algpseudocode} 
\usepackage{array} 
\usepackage{enumitem}
\usepackage[caption=false,font=footnotesize,labelfont=sf,textfont=sf]{subfig}
\usepackage{textcomp}
\usepackage{stfloats}
\usepackage{booktabs}
\usepackage{url}
\usepackage{verbatim}

\usepackage[dvipsnames]{xcolor}
\usepackage{diagbox}
\usepackage{multirow}
\usepackage[T1]{fontenc}
\usepackage{textcomp}
\usepackage{xcolor}
\usepackage{bm}

\def\BibTeX{{\rm B\kern-.05em{\sc i\kern-.025em b}\kern-.08em
    T\kern-.1667em\lower.7ex\hbox{E}\kern-.125emX}}

\begin{document}

\title{From Pilots to Position: Experimental Near-Field Localization for ISAC
\thanks{
This work is supported by the SUNRISE-6G project,
funded by the Smart Networks and Services Joint Undertaking (SNS JU)
under the European Union’s Horizon Europe research and innovation programme, with Grant Agreement No. 101139257. The work of Rodney Martinez Alonso is supported by the Research Foundation–Flanders (FWO) under Grant 1211926N.}}

\author{
\IEEEauthorblockN{
Yunxin Li\IEEEauthorrefmark{1}, 
Emiel Vanspranghels\IEEEauthorrefmark{1}\IEEEauthorrefmark{2}, 
Jialun Kou\IEEEauthorrefmark{1}\IEEEauthorrefmark{2}, 
Georgios Mystriotis\IEEEauthorrefmark{1},\\
Achiel Colpaert\IEEEauthorrefmark{2}\IEEEauthorrefmark{1}, 
Rodney Martinez Alonso\IEEEauthorrefmark{1}\IEEEauthorrefmark{2}, 
Sofie Pollin\IEEEauthorrefmark{1}\IEEEauthorrefmark{2}}

\IEEEauthorblockA{\IEEEauthorrefmark{1}Department of Electrical Engineering (ESAT), KU Leuven, 3001 Leuven, Belgium}
\IEEEauthorblockA{\IEEEauthorrefmark{2}%
imec, Kapeldreef 75, 3001 Leuven, Belgium}
\IEEEauthorblockA{Email: \{\textit{firstname.lastname}\}\textit{@kuleuven.be}}
}

\maketitle

\begin{abstract}

Integrated sensing and communication (ISAC) enables wireless systems to reuse communication signals for environmental sensing, potentially reducing the need for dedicated sensing resources. However, the achievable localization accuracy of communication-centric waveforms under practical hardware constraints remains an important challenge. In this paper, we experimentally investigate near-field target localization using OFDM pilot signals in a distributed communication system. We consider both coherent and noncoherent localization approaches, exploiting multistatic propagation delays and time-difference-of-arrival (TDOA) information extracted from pilot-based channel estimates. A hardware setup comprising two transmitters and four receivers was developed and configured to operate at 3.4 GHz with 198 MHz occupied bandwidth to evaluate localization performance under realistic propagation and synchronization conditions. Experimental results across 16 static target positions demonstrate mean localization errors of 19.8 cm and 18.8 cm for coherent backprojection and noncoherent delay–TDOA fusion, respectively. Furthermore, we investigate the impact of occupied bandwidth on localization accuracy of the two proposed algorithms. These findings provide experimental insights into the capabilities and practical limitations of OFDM pilot-based sensing, supporting resource-efficient localization in future ISAC systems. 
\end{abstract}


\section{Introduction}
\label{sec:introduction}

Integrated sensing and communication (ISAC) is emerging as a key capability of future wireless networks, enabling communication infrastructure to provide environmental awareness alongside conventional data transmission. By sharing spectrum, hardware, and signal resources, ISAC offers the potential to support sensing functionalities without deploying dedicated radar systems~\cite{liu2022isac}. Among its applications, passive target localization is particularly attractive for indoor automation, robotics, and intelligent wireless environments, where the positions of objects without active communication capabilities need to be determined. In particular, known pilot symbols allow the extraction of channel frequency responses (CFRs), from which target-induced propagation changes can be exploited for localization. This creates opportunities to extend the functionality of communication signals toward environmental sensing~\cite{liu2022isac}. Despite this potential, accurate near-field indoor localization using communication waveforms remains challenging due to their finite bandwidth and practical hardware constraints, where target reflections often overlap with direct-path components and environmental multipath. Nevertheless, spatially separated transmitters and receivers observe different bistatic propagation paths, providing complementary geometric constraints on the target position. By combining these spatial observations, localization can exploit information beyond the delay resolution of any individual link. Therefore, understanding how frequency-domain information and spatial diversity jointly contribute to localization performance is essential for practical ISAC system design~\cite{sakhnini2022coherent}.

Recent research has extensively investigated the theoretical foundations of near-field sensing and localization. Near-field ISAC frameworks have explored joint distance and angle estimation through Cram\'{e}r--Rao bound (CRB) analysis and beamforming optimization~\cite{li2024nearfield}, while cooperative and multistatic localization schemes using 5G NR OFDM signals and communication preambles have further explored TDOA, spatial backprojection, and fusion of measurements across multiple access points, with performance primarily assessed through numerical simulations~\cite{zhang2025cooperative,tosun2026prach}. Experimental ISAC prototypes have also demonstrated promising localization capabilities under different hardware and signal configurations. Near-field coherent sensing using an OFDM massive-MIMO communication testbed has achieved centimeter-level localization accuracy despite an occupied bandwidth of only 18~MHz, highlighting the substantial spatial information provided by large antenna arrays~\cite{sakhnini2022coherent,sakhnini2022robust}. More recently, a bistatic MIMO-OFDM prototype operating at 28~GHz with 1.92~GHz bandwidth has demonstrated real-time target localization and tracking~\cite{dubs2026realtime}. Meanwhile, hardware measurements have shown that phase inconsistencies across OFDM channels and antennas can severely degrade localization accuracy, motivating dedicated calibration procedures for coherent processing~\cite{zhang2026phase}. Collectively, these studies demonstrate the feasibility of OFDM-based sensing, but also highlight the dependence of localization performance on available bandwidth, spatial aperture, and hardware coherence. Comparatively less experimental evidence is available on the accuracy achievable with a small number of distributed sensing channels and moderate occupied bandwidth.

To bridge this research gap, we investigate near-field passive target localization using communication OFDM pilots in a distributed multistatic testbed operating at 3.4~GHz with 198~MHz occupied bandwidth. Rather than relying on large antenna arrays or GHz-level sensing bandwidth, we focus on exploiting multistatic geometry and complementary delay information from a limited number of links. The proposed localization method achieves a mean positioning error of 18.8~cm and an RMSE of 20.3~cm, despite the nominal bistatic path length resolution of approximately 1.51~m. 

\section{System Model}
\label{sec:system_model}

\begin{figure}[t]
    \centering
    \includegraphics[width=\linewidth]{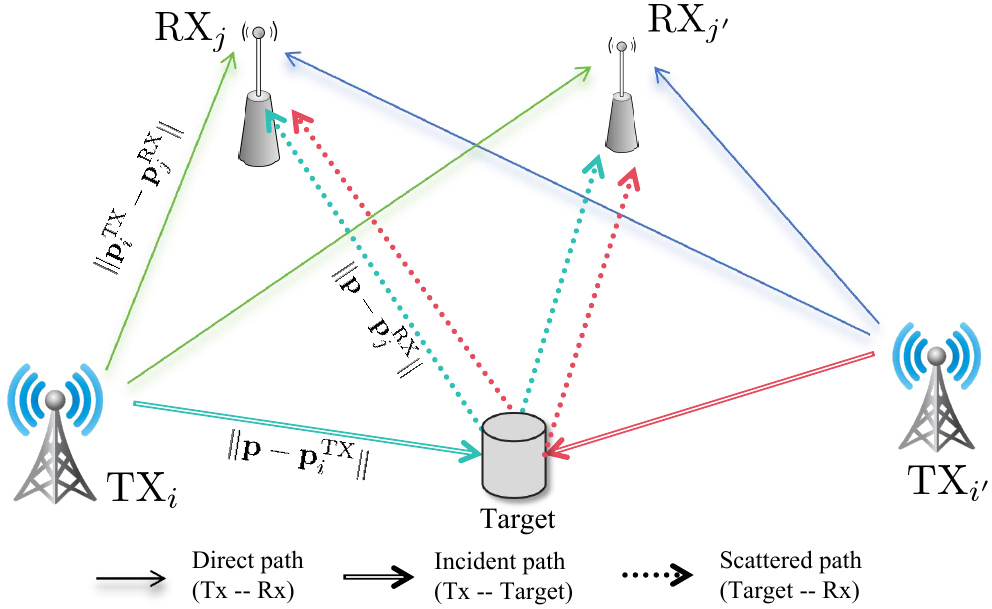}
    \caption{Geometry illustration of the considered multistatic ISAC system.}
    \label{fig:system}
\end{figure}

We consider a distributed OFDM-based ISAC system
consisting of $N_{\mathrm T}$ transmitters (TXs) and
$N_{\mathrm R}$ receivers (RXs), as illustrated in
Fig.~\ref{fig:system}. The TXs and RXs are deployed
at known locations to localize a stationary target, modeled as an isotropic point scatterer,
within a two-dimensional region of interest (RoI). 
The system employs communication-centric OFDM
pilot signals, with the TXs operating in a
time-division multiplexing (TDM) manner. During
each transmission interval, only one TX is active,
while all RXs collect the corresponding signals.

\subsection{OFDM Signal and Channel Model}

We consider an OFDM waveform with $M$ occupied
subcarriers and subcarrier spacing $\Delta f$.
Let $L$ denote the number of observed OFDM symbols
within a frame.
The frequency domain
signal transmitted by TX $i$ is represented by
the OFDM resource grid
$\mathbf{X}_i\in\mathbb{C}^{M\times L}$,
where each row corresponds to a subcarrier and
each column corresponds to an OFDM symbol.
The transmitted pilot symbols are known at the receiver side and are used for channel estimation. 

After FFT processing, the received signal at
RX $j$ from TX $i$ is expressed as
\begin{equation}
\mathbf{Y}_{i,j}
=
\mathbf{H}_{i,j}\odot\mathbf{X}_i
+
\mathbf{Z}_{i,j},
\label{eq:received_signal}
\end{equation}
where $\mathbf{H}_{i,j}\in\mathbb{C}^{M\times L}$
denotes the channel frequency response (CFR),
$\mathbf{Z}_{i,j}\in\mathbb{C}^{M\times L}$
represents additive noise, and $\odot$ denotes
the Hadamard product. The received signal
$\mathbf{Y}_{i,j}$ has the same dimensions as
the transmitted resource grid.

Since the transmitted symbols are known and
nonzero on the occupied resource elements,
the CFR can be estimated as
\begin{equation}
\widehat{\mathbf{H}}_{i,j}
=
\left[
\frac{Y_{i,j}[m,\ell]}
     {X_i[m,\ell]}
\right]_{m,\ell},
\label{eq:cfr_estimation}
\end{equation}
where $m=1,\ldots,M$ and $\ell=1,\ldots,L$.

To distinguish target-induced propagation
changes from the static environment,
measurements are collected both with and
without the target. These measurements serve as the inputs to the processing explained in Section~\ref{sec:localization}. 

\subsection{Near-Field Bistatic Propagation Model}

Let $\mathbf{p}\in\mathbb{R}^{2}$ denote the
unknown target position, and let
$\mathbf{p}_{i}^{\mathrm{TX}}\in\mathbb{R}^{2}$
and $\mathbf{p}_{j}^{\mathrm{RX}}\in\mathbb{R}^{2}$
denote the known positions of TX $i$ and RX $j$,
respectively. Under a point-scatterer approximation,
the bistatic propagation distance through the
target is given by
\begin{equation}
d_{i,j}(\mathbf{p})
=
\|\mathbf{p}-\mathbf{p}_{i}^{\mathrm{TX}}\|
+
\|\mathbf{p}-\mathbf{p}_{j}^{\mathrm{RX}}\|,
\label{eq:bistatic_distance}
\end{equation}
where $\|\cdot\|$ denotes the Euclidean norm.
The corresponding bistatic propagation delay is
$\tau_{i,j}(\mathbf{p})=d_{i,j}(\mathbf{p})/c$,
where $c$ is the speed of light.

Under a simplified single scatterer model,
the target-reflected channel component on
subcarrier $m$ can be expressed as
\begin{equation}
h_{i,j}[m]
=
\alpha_{i,j}
e^{-j2\pi(f_c+f_m)\tau_{i,j}(\mathbf{p})},
\label{eq:target_channel}
\end{equation} 
where $f_c$ denotes the carrier frequency,
$f_m$ is the baseband frequency of the
$m$-th occupied subcarrier, and
$\alpha_{i,j}$ represents the complex
scattering and propagation coefficient
associated with the TX--target--RX path. 
The coefficient may differ across links,
accounting for their distinct propagation
conditions.

Using the direct TX--RX path as a reference,
the bistatic excess path length is defined as
\begin{equation}
e_{i,j}(\mathbf{p})
=
d_{i,j}(\mathbf{p})
-
\|\mathbf{p}_{i}^{\mathrm{TX}}
-\mathbf{p}_{j}^{\mathrm{RX}}\|.
\label{eq:excess_path}
\end{equation}
The corresponding excess propagation delay is
\begin{equation}
\tau_{i,j}^{\mathrm{ex}}(\mathbf{p})
=
\frac{e_{i,j}(\mathbf{p})}{c}.
\label{eq:excess_delay}
\end{equation}

Furthermore, for two receivers $j$ and $j'$
observing the same transmitter $i$, the
difference between their excess delays is
\begin{equation}
\begin{aligned}
\Delta\tau_{i,j,j'}(\mathbf{p})
&=
\tau_{i,j}^{\mathrm{ex}}(\mathbf{p})
-
\tau_{i,j'}^{\mathrm{ex}}(\mathbf{p})\\
&=
\frac{1}{c}\Big(
\|\mathbf{p}-\mathbf{p}_{j}^{\mathrm{RX}}\|
-
\|\mathbf{p}-\mathbf{p}_{j'}^{\mathrm{RX}}\|\\
&\quad-
\|\mathbf{p}_{i}^{\mathrm{TX}}
-\mathbf{p}_{j}^{\mathrm{RX}}\|
+
\|\mathbf{p}_{i}^{\mathrm{TX}}
-\mathbf{p}_{j'}^{\mathrm{RX}}\|
\Big).
\end{aligned}
\label{eq:tdoa_geometry}
\end{equation}
The TX--target propagation distance cancels
in the receiver-pair difference, providing
an additional geometric constraint on the
target position.

The bistatic propagation delays and
receiver-pair delay differences form the
geometric basis for the coherent backprojection
and noncoherent delay--TDOA localization methods,
respectively, as detailed in
Section~\ref{sec:localization}.

\section{OFDM Pilot-Based Localization Methods}
\label{sec:localization}

This section presents two localization methods
based on the pilot-derived CFRs and bistatic
propagation model introduced in
Section~\ref{sec:system_model}. Both methods
exploit target-induced channel variations
to construct spatial localization maps.
Coherent backprojection (BP) combines
phase-compensated channel responses, whereas
noncoherent delay--TDOA fusion integrates
bistatic excess delay and receiver-pair TDOA evidence.

\subsection{CFR Processing and Delay-Domain Representation}
\label{subsec:channel_processing}

For each TX--RX link, measurements collected
with and without the target are processed
separately. Let
$\widehat{\mathbf H}_{i,j}^{a,n}
\in\mathbb C^{M\times L}$
denote the estimated CFR matrix of frame $n$,
where $a\in\{\mathrm{tar},\mathrm{bg}\}$
identifies the measurement scene, and
$n=1,\ldots,N$ indexes the selected frames.

Both methods employ temporal accumulation
and background subtraction, but use different
calibration procedures. In coherent BP, complex CFRs
are averaged directly without additional
fitted phase alignment. In delay--TDOA fusion,
the CFRs undergo symbol-level common phase
error correction and frame-level phase/timing
alignment. A dominant background path is
further used for link-specific delay and
complex gain calibration.

Let
$\widetilde{\mathbf H}_{i,j}^{a,n,\nu}
\in\mathbb C^{M\times L}$
denote the CFR matrix after the corresponding
processing, where $\nu\in\{\mathrm C,\mathrm D\}$
identifies coherent BP and delay--TDOA, respectively.
For coherent BP,
$\widetilde{\mathbf H}_{i,j}^{a,n,\mathrm C}
=\widehat{\mathbf H}_{i,j}^{a,n}$.
The temporally averaged CFR is expressed as
\begin{equation}
\widetilde{\mathbf h}_{i,j}^{a,\nu}
=
\frac{1}{NL}
\sum_{n=1}^{N}
\widetilde{\mathbf H}_{i,j}^{a,n,\nu}
\mathbf1_L,
\label{eq:temporal_processing}
\end{equation}
where $\mathbf1_L$ denotes an all-one vector
of length $L$.

To suppress static environmental responses,
the target and background CFRs are subtracted:
\begin{equation}
\Delta\mathbf h_{i,j}^{\nu}
=
\widetilde{\mathbf h}_{i,j}^{\mathrm{tar},\nu}
-
\widetilde{\mathbf h}_{i,j}^{\mathrm{bg},\nu}.
\label{eq:background_subtraction}
\end{equation}
The residual CFR is transformed into the
complex delay domain through
\begin{equation}
g_{i,j}^{\nu}(\tau)
=
\frac{1}{N_{\mathrm{FFT}}}
\sum_{m=1}^{M}
\Delta h_{i,j}^{\nu}[m]
e^{j2\pi f_m\tau},
\label{eq:delay_transform}
\end{equation}
where $N_{\mathrm{FFT}}$ is the OFDM FFT size,
$f_m$ denotes the baseband frequency of
subcarrier $m$, and $\tau$ is the delay
variable. The occupied subcarriers are
mapped to their original frequency bins
before inverse Fourier transformation.
Zero-padding provides finer delay domain
sampling without increasing the intrinsic
delay resolution.

The complex delay responses form the basis
of both localization methods. Their delay
references and spatial combination rules
are described in the following subsections.
Both methods construct spatial scores over
a two-dimensional candidate grid
$\mathcal P$. The target position is
estimated by
\begin{equation}
\widehat{\mathbf p}_{\nu}
=
\underset{\mathbf p\in\mathcal P}{\arg\max}
S_{\nu}(\mathbf p),
\qquad
\nu\in\{\mathrm C,\mathrm D\},
\label{eq:position_estimation}
\end{equation}
where the corresponding spatial scores
are defined in the following subsections.

\subsection{Coherent Backprojection}
\label{subsec:coherent_bp}

Coherent BP exploits the bistatic propagation
geometry to focus complex channel responses
onto target positions. For each
TX--RX pair, a measured propagation delay
constrains the target to an ellipse whose
foci are the corresponding TX and RX
positions. Combining multiple link responses
enables spatial localization, while coherent
summation additionally exploits their
relative phases.

For position $\mathbf p$, the
expected bistatic delay
$\tau_{i,j}(\mathbf p)$ follows from
\eqref{eq:bistatic_distance}. In practice,
hardware-dependent timing offsets cause
the observed CIRs to deviate from their
geometrically predicted delays. These offsets
are estimated from the dominant background
responses relative to the known direct
TX--RX propagation delays.

The phase-compensated
backprojection contribution of link $(i,j)$
is expressed as 
\begin{equation}
B_{i,j}(\mathbf p)
={}
g_{i,j}^{\mathrm C}
\left( \tau_{i,j}(\mathbf p) \right)
e^{j2\pi f_c\tau_{i,j}(\mathbf p)}. 
\label{eq:cbp_contribution}
\end{equation}

The complex delay response is evaluated
at the timing-corrected bistatic delay,
while the exponential compensates for
the propagation-dependent carrier phase.

The contributions from all TX--RX links
are coherently combined to construct
the spatial power map:
\begin{equation}
S_{\mathrm C}(\mathbf p)
=
\left|
\sum_{i=1}^{N_{\mathrm T}}
\sum_{j=1}^{N_{\mathrm R}}
B_{i,j}(\mathbf p)
\right|^2.
\label{eq:cbp_map}
\end{equation}

At the target position, consistent delay
and phase compensation can produce
constructive summation across links.
However, the effectiveness of coherent
focusing depends on inter-link phase
consistency, which may be affected by
residual RF and cable phase offsets.

\subsection{Noncoherent Delay--TDOA Fusion}
\label{subsec:delay_tdoa}

Unlike coherent BP, the delay--TDOA method combines
nonnegative spatial evidence without requiring
inter-link carrier-phase coherence.
It exploits two complementary geometric
constraints: bistatic excess delays from
individual TX--RX links and differential
excess delays between receiver pairs.

For this method, the dominant background
path serves as a link-specific delay
reference. Its estimated delay
$\widehat\tau_{i,j}^{\mathrm{ref}}$
and complex gain
$\widehat\alpha_{i,j}^{\mathrm{ref}}$
are obtained from the peak of the background
delay response and its corresponding complex
amplitude, respectively. The CFR calibration
applies the frequency-dependent factor
$\gamma_{i,j}[m]
=e^{j2\pi f_m\widehat\tau_{i,j}^{\mathrm{ref}}}
/\widehat\alpha_{i,j}^{\mathrm{ref}}$
to both target and background measurements.
This references the residual CFR to the
dominant background path, being the direct path between TX $i$ and RX $j$, allowing delay
processing in terms of excess propagation
delays. 

\subsubsection{Bistatic Delay Evidence}

For each TX--RX link, the magnitude-squared
delay response provides evidence of
target-induced reflections. In the
experimental implementation, the transform
in \eqref{eq:delay_transform} is evaluated
with variance-based frequency weighting.
Noise-bias correction and per-link
normalization are subsequently applied
to obtain a nonnegative delay-evidence
profile $P_{i,j}(\tau)$.

For position $\mathbf p$,
each link contributes evidence at the
predicted excess bistatic delay
$\tau_{i,j}^{\mathrm{ex}}(\mathbf p)$
defined in \eqref{eq:excess_delay}.
The corresponding spatial delay score
is obtained by accumulating evidence
from multiple bistatic links. 

\subsubsection{Receiver-Pair TDOA Evidence}

Additional spatial information is obtained
from pairs of receivers observing the
same transmitter. For RXs $j$ and $j'$,
the cross-spectrum of their residual CFRs
is defined as
\begin{equation}
\mathbf c_{i,j,j'}
=
\Delta\mathbf h_{i,j}^{\mathrm D}
\odot
\left(
\Delta\mathbf h_{i,j'}^{\mathrm D}
\right)^*,
\label{eq:tdoa_cross}
\end{equation}
where $(\cdot)^*$ denotes element-wise
complex conjugation.

Applying the frequency-to-delay
transformation to the cross-spectrum gives
the differential-delay response:
\begin{equation}
q_{i,j,j'}(\tau)
=
\frac{1}{N_{\mathrm{FFT}}}
\sum_{m=1}^{M}
c_{i,j,j'}[m]e^{j2\pi f_m\tau}.
\label{eq:tdoa_response}
\end{equation}
A frequency-weighted version of this
response is used to construct the
normalized differential-delay power
profile $Q_{i,j,j'}(\tau)$.

For a position $\mathbf p$,
the predicted differential excess delay
$\Delta\tau_{i,j,j'}(\mathbf p)$ is
defined in \eqref{eq:tdoa_geometry}.
Since the two receivers share the same
transmitter, the TX--target propagation
distance cancels in their delay difference.
This provides additional geometric
information complementary to the
individual bistatic delays.

\subsubsection{Spatial Score Fusion}

The delay and TDOA profiles are evaluated
at the predicted delays of each
position $\mathbf p$, yielding two spatial
scores:
\begin{equation}
\begin{aligned}
D(\mathbf p)
&=
\sum_{i=1}^{N_{\mathrm T}}
\sum_{j=1}^{N_{\mathrm R}}
\lambda_{i,j}
P_{i,j}\left(
\tau_{i,j}^{\mathrm{ex}}(\mathbf p)
\right),\\
T(\mathbf p)
&=
\sum_{i=1}^{N_{\mathrm T}}
\sum_{j<j'}
\sqrt{\lambda_{i,j}\lambda_{i,j'}}
Q_{i,j,j'}\left(
\Delta\tau_{i,j,j'}(\mathbf p)
\right),
\end{aligned}
\label{eq:delay_tdoa_maps}
\end{equation}
where $\lambda_{i,j}\in[0,1]$ denotes
the reliability weight of link $(i,j)$.
The geometric mean
$\sqrt{\lambda_{i,j}\lambda_{i,j'}}$
weights each TDOA score according to the
reliabilities of its two associated links.

The link weights are determined from
the target-induced channel change
relative to the estimated noise level:
\begin{equation}
\begin{aligned}
\chi_{i,j}
&=
\frac{
\sum_m w_{i,j}[m]
|\Delta h_{i,j}^{\mathrm D}[m]|^2
}{
\sum_m w_{i,j}[m]\xi_{i,j}[m]
},\\
\lambda_{i,j}
&=
\left[
\frac{\chi_{i,j}-1}{4}
\right]_0^1,
\end{aligned}
\label{eq:link_reliability}
\end{equation}
where $w_{i,j}[m]$ is the
frequency-domain inverse-variance weight,
$\xi_{i,j}[m]$ is the estimated variance
of the target-minus-background mean CFR,
and $[x]_0^1=\min\{1,\max\{0,x\}\}$.
The mapping from $\chi_{i,j}$ to
$\lambda_{i,j}$ provides a bounded,
heuristic measure of link reliability.

Finally, the two spatial scores are
normalized and combined as
\begin{equation}
S_{\mathrm D}(\mathbf p)
=
\frac{D(\mathbf p)}
{\max_{\mathbf q\in\mathcal P}D(\mathbf q)}
\frac{T(\mathbf p)}
{\max_{\mathbf q\in\mathcal P}T(\mathbf q)}.
\label{eq:delay_tdoa_fusion}
\end{equation}
This multiplicative fusion favors positions supported by
both bistatic delay and TDOA evidence.

\section{Experimental Setup}
\begin{figure}[t]
    \centering
    \includegraphics[width=\linewidth]{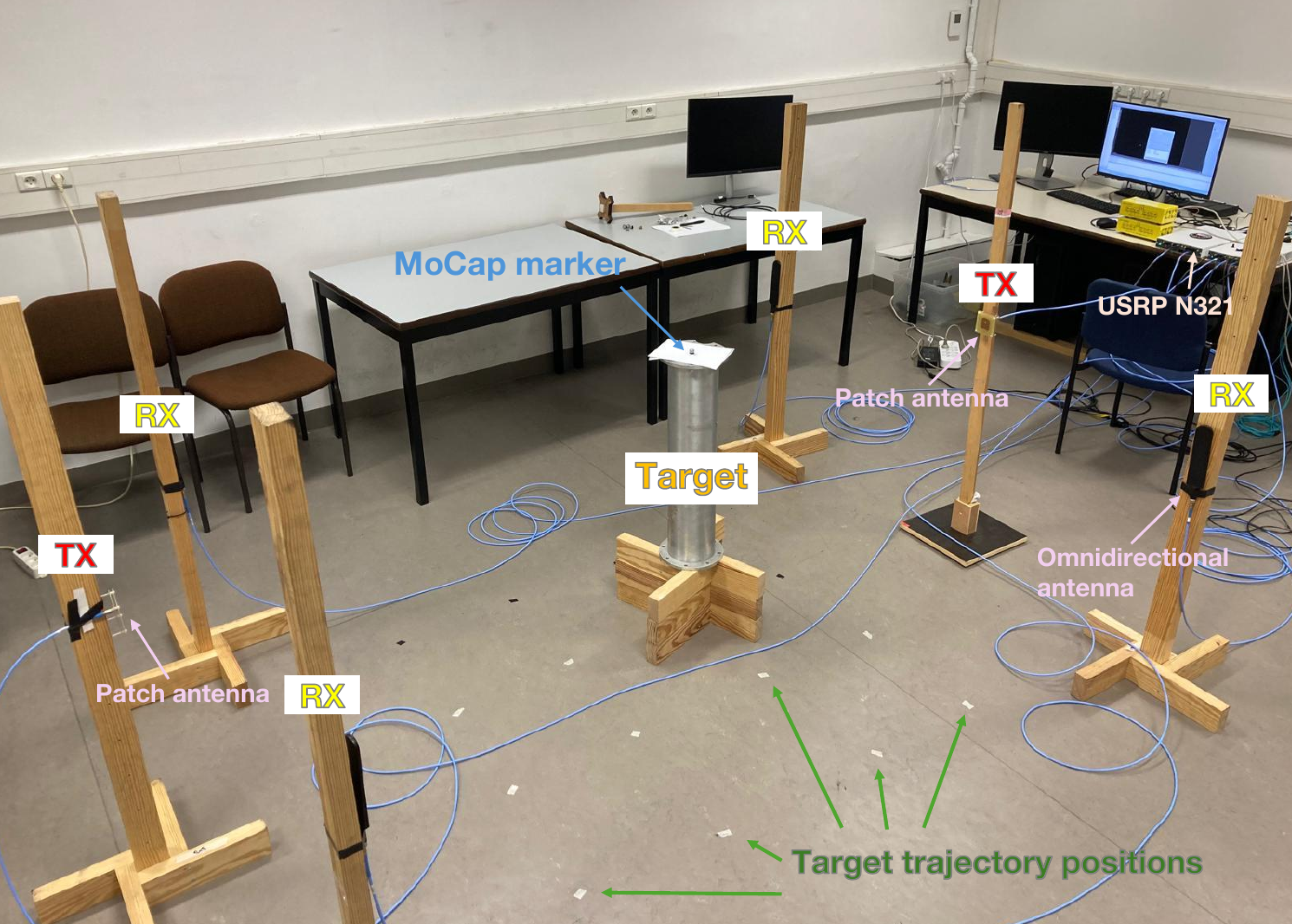}
    \caption{Laboratory measurement setup for OFDM near-field multistatic passive target localization.}
    \label{fig:labsetup}
\end{figure}

This section describes the experimental setup used to collect channel measurements and ground-truth positions, including the RF hardware, synchronization, and position labeling procedure. The laboratory measurement setup is shown in Fig.~\ref{fig:labsetup}.

\subsection{RF Hardware}
\begin{itemize}
    \item \textbf{Digital baseband and FR1 SDR -- USRP N321:}
    The Ettus Research USRP N321 performs digital baseband processing, digital-to-analog and analog-to-digital conversion, and RF up/downconversion. It supports RF frequencies from 3 MHz to 6 GHz. In our measurements, the sampling rate was set to 245.76 MS/s\footnote{This sampling rate supports power-of-two FFT sizes at standard OFDM subcarrier spacings, facilitating efficient FFT processing.}.

    \item \textbf{Synchronization -- NI CDA-2990 and LO sharing:}
    An NI CDA-2990 distributes a common pulse-per-second (PPS) signal and a 10 MHz reference clock to the USRP N321 devices for time and frequency synchronization, respectively. Additionally, local oscillator (LO) sharing between the USRP channels enables phase-coherent operation.

    \item \textbf{Antennas:}
    The four RXs were equipped with omnidirectional antennas to capture multipath components from a wide range of directions, while the two TXs used patch antennas \cite{patch_Vladi} to provide higher gain toward the RoI.
    
\end{itemize}

\subsection{Position Labeling}

In the experiments,
a metallic cylinder is used as the target,
whose actual scattering response may vary
with the bistatic observation geometry. Ground-truth positions of the target, TXs, and RXs were obtained using a Qualisys Miqus marker-based motion-capture (MoCap) system. A single Miqus M3 camera was mounted above the setup with its optical axis perpendicular to the floor, covering the RoI. Since only planar coordinates were required, tracking was restricted to the horizontal plane.

The TX and RX positions were recorded before the measurements and used as known inputs to the algorithms described in this paper. The target positions were recorded concurrently with the RF measurements and used as ground truth to evaluate the localization performance.

\section{Experimental Results}
\label{sec:experimental_results}

We evaluate coherent BP and noncoherent
joint delay--TDOA localization
using indoor OFDM measurements at a carrier frequency of
3.4~GHz. The bandwidth is varied from
39.6 to 198~MHz, with other experimental parameters listed in Table~\ref{tab:Sim}. The MoCap system provides
reference coordinates at 16 stationary target locations as ground truth. For the bandwidth comparison, a common set of 20 background frames is shared across all positions, while 20 frames with the target present are used at each of the positions. Both localization methods are evaluated using
the same recordings at each bandwidth setting,
ensuring identical measurement conditions
for the comparison.

\begin{table}[!h]
    \caption{Experimental Parameters} \label{tab:Sim} 
    \normalsize
    \centering
    \begin{tabular}{p{1.7cm} p{1.7cm} p{1.7cm} p{1.7cm}} 
    \hline
    \hline
    Parameter & Value & Parameter & Value\\
    \hline
    $f_c$ & $3.4\operatorname{GHz}$ & $f_s$ & $245.76\operatorname{MSps}$\\
    $B$ & $198\operatorname{MHz}$ & $\Delta f$ & $30\operatorname{kHz}$\\
    $N_\text{T}$ & $2$ & $N_\text{R}$ & $4$\\
    $N_\text{FFT}$ & $8192$ & $M$ & $6600$\\
    $N$ & $20$ & $L$ & $64$\\
    \hline
    \end{tabular}
\end{table}

For target position $k$, the two-dimensional Euclidean
localization error is defined as
\begin{equation}
    e_k = \left\|\widehat{\mathbf p}_k
    - \mathbf p_k^{\mathrm{GT}}\right\|_2,
\end{equation}
where $\widehat{\mathbf p}_k$ and
$\mathbf p_k^{\mathrm{GT}}$ are the estimated and MoCap
reference coordinates, respectively. We report the mean
error $\bar e=K^{-1}\sum_{k=1}^K e_k$, the root mean square
error (RMSE) $\sqrt{K^{-1}\sum_{k=1}^K e_k^2}$, and the
median error, with $K=16$.

\subsection{Localization Accuracy}
\label{subsec:localization_accuracy}

\begin{figure*}[t]
    \centering
    \subfloat[Coherent BP.]{%
        \includegraphics[width=0.47\textwidth]{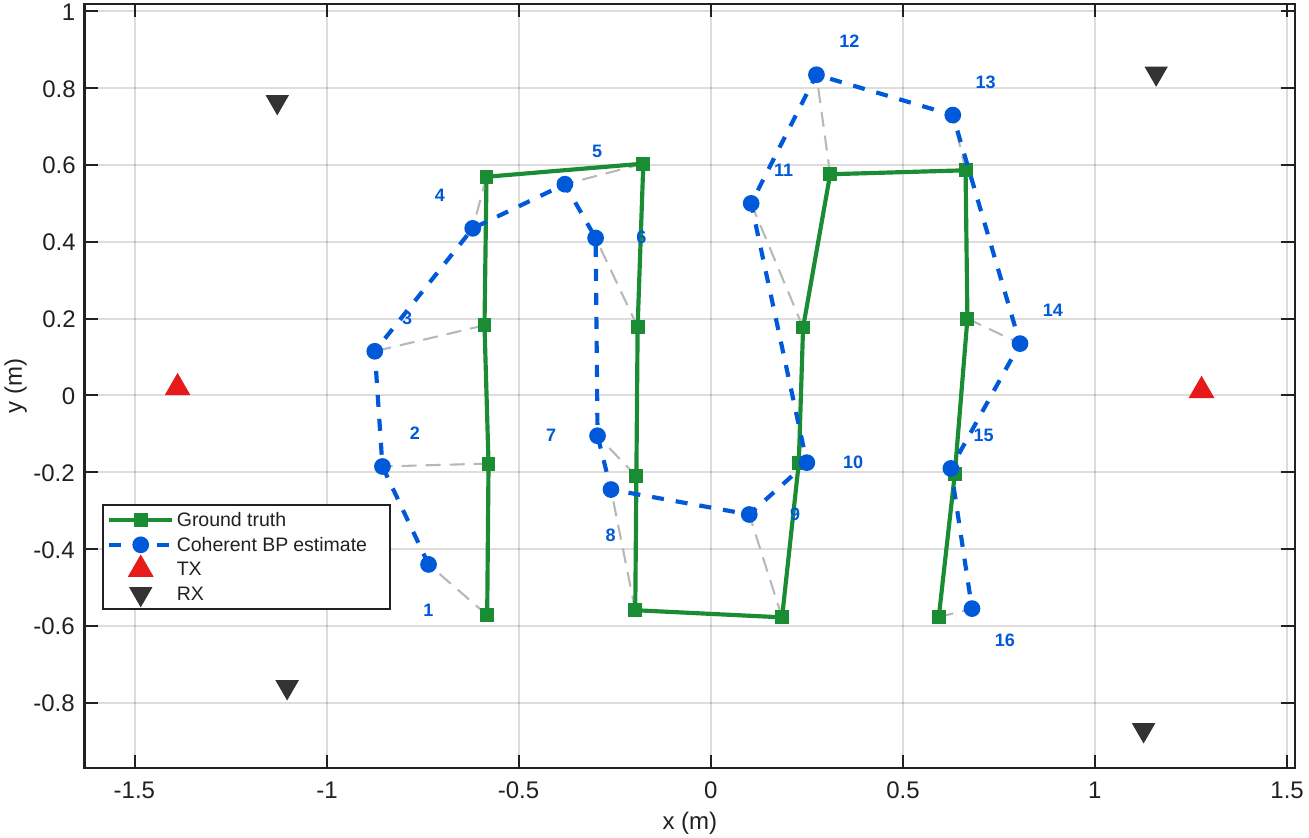}%
        \label{fig:localization_bp}}
    \hfill
    \subfloat[Noncoherent delay--TDOA fusion.]{%
        \includegraphics[width=0.47\textwidth]{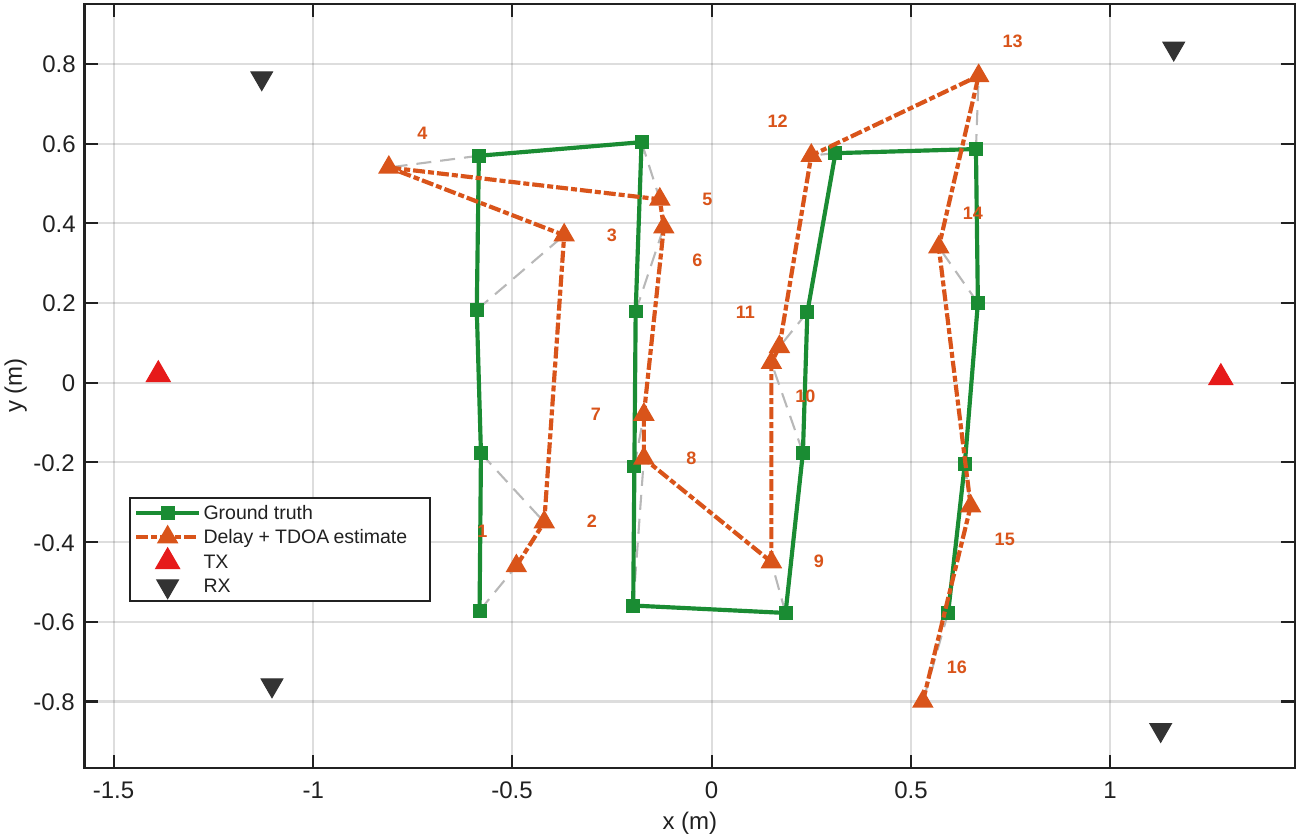}%
        \label{fig:localization_dt}}
    \caption{Localization results at 16 stationary
    target positions using (a) coherent BP and
    (b) non-coherent delay--TDOA fusion.
    Gray dashed segments connect corresponding
    estimates and MoCap ground-truth positions.
    Colored lines connect consecutive indexed
    measurement positions for visualization only.}
    \label{fig:localization_positions}
\end{figure*}

Fig.~\ref{fig:localization_positions} presents the
localization results. Both coherent
BP and noncoherent delay--TDOA fusion successfully
localize the target across the measurement region,
although position-dependent deviations are observed.

An important observation is that both methods achieve
decimeter-level localization accuracy despite the
limited signal bandwidth. At the maximum occupied
bandwidth of $B=198$~MHz, the nominal bistatic
path-length resolution of an individual sensing link is
$\Delta L_{\mathrm{nom}}\approx \frac{c}{B}\approx 1.51~\mathrm{m}$, where $c$ denotes the speed of light. In the considered distributed configuration, the
two TXs and four RXs provide multiple
bistatic observations with different spatial
geometries. Each observation constrains the target
to a set of positions consistent with its
propagation path length, while the combination of
multiple links helps resolve spatial ambiguities.
Both localization methods exploit this geometric
diversity to identify a common target position, therefore, sub-meter localization accuracy can
be achieved.

\begin{table}[t]
\centering
\caption{Localization accuracy at 198~MHz
across 16 positions (mm)}
\label{tab:accuracy_results}
\footnotesize
\setlength{\tabcolsep}{3pt}
\begin{tabular}{lrrrr}
\toprule
Method & Mean & RMSE & Median & Max. \\
\midrule
Coherent BP
    & 197.8 & 220.9 & 206.5 & \textbf{350.3} \\
Delay only
    & 203.6 & 227.7 & 177.7 & 466.1 \\
TDOA only
    & 237.0 & 270.8 & 216.3 & 594.5 \\
Delay--TDOA fusion
    & \textbf{188.1} & \textbf{202.7}
    & \textbf{177.4} & 370.2 \\
\bottomrule
\end{tabular}
\end{table}

As summarized in Table~\ref{tab:accuracy_results},
coherent BP achieves a localization RMSE of 220.9~mm, while
non-coherent delay--TDOA fusion achieves 202.7~mm.
The latter reduces the RMSE by 11.0\% relative
to delay-only localization and by 25.2\% relative
to TDOA-only localization. These improvements indicate that combining absolute
bistatic delay constraints with relative delay
constraints provides a more discriminative spatial
localization score than either component alone
in the considered environment. More importantly,
the results demonstrate that spatial fusion across
distributed sensing links can provide localization
accuracy substantially finer than the nominal
single-link path-length resolution.

\subsection{Impact of Signal Bandwidth}
\label{subsec:bandwidth_results}

\begin{figure}[t]
    \centering
    \includegraphics[width=0.8\linewidth]{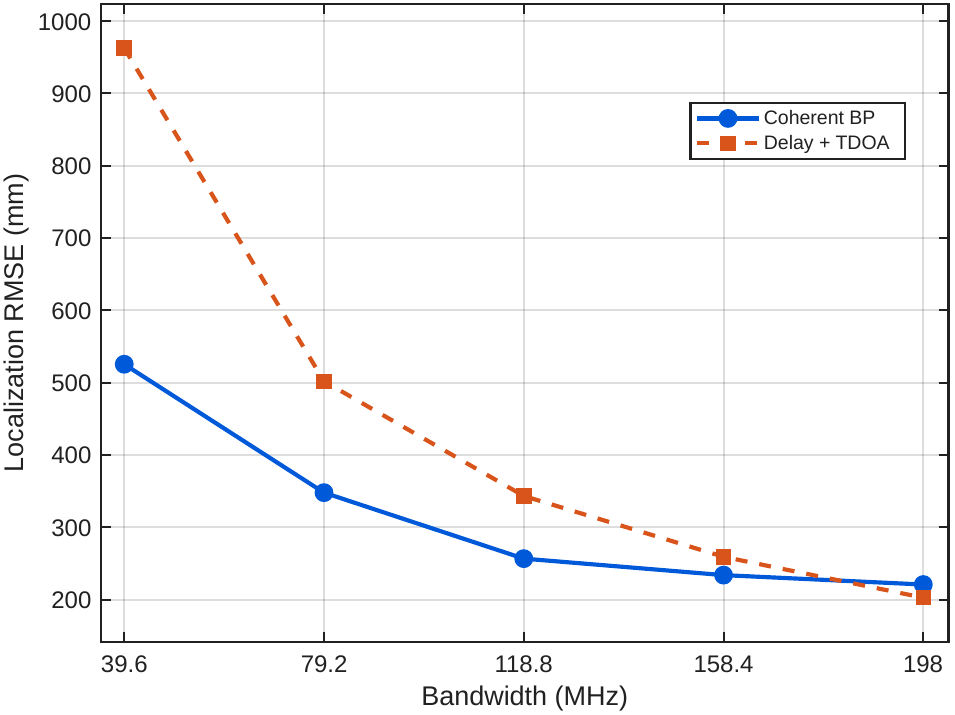}
    \caption{Localization RMSE versus occupied
    signal bandwidth for coherent BP and
    noncoherent delay--TDOA fusion.
    All configurations use 16 target positions, with 20 target frames per position and a common set of 20 background frames.}
    \label{fig:bandwidth_results}
\end{figure}

To investigate the role of signal bandwidth,
Fig.~\ref{fig:bandwidth_results} compares the
localization RMSE of the two methods at five
bandwidth configurations, ranging from
39.6 to 198~MHz.

Both methods benefit from increasing bandwidth,
but exhibit different levels of sensitivity.
At 39.6~MHz, coherent BP achieves an RMSE
of 525.4~mm, compared with 962.5~mm for
delay--TDOA fusion. This advantage can be
attributed in part to their different
information utilization strategies.
With limited bandwidth, delay and TDOA
profiles provide relatively coarse propagation
delay discrimination, which can result in
broad or ambiguous spatial constraints.
In contrast, coherent BP operates directly
on the complex channel responses, retaining
phase-dependent information during spatial
processing. Such information can help
distinguish target positions when
delay-based constraints are less informative.

As bandwidth increases, delay discrimination
improves, providing more concentrated spatial
constraints for non-coherent fusion.
Accordingly, delay--TDOA fusion exhibits
a larger relative improvement, reducing
its RMSE by 78.9\% from 39.6 to 198~MHz,
compared with a 58.0\% reduction for
coherent BP. At 198~MHz, delay--TDOA fusion
slightly outperforms coherent BP,
achieving an 8.2\% lower RMSE.

These results reveal a practical tradeoff between the two localization approaches. Coherent BP achieves lower localization errors under reduced bandwidth, but requires phase-consistent channel measurements for coherent spatial combining. In contrast, noncoherent delay--TDOA fusion avoids cross-link phase-coherent combining, while being more sensitive to bandwidth limitations.
As bandwidth increases, the improved delay
discrimination enables delay--TDOA fusion
to achieve a lower localization error than
coherent BP at 198~MHz.
Despite their different processing requirements and bandwidth sensitivities, both approaches achieve approximately 20~cm mean localization error with 198~MHz of occupied bandwidth,
demonstrating the potential of distributed
OFDM sensing for accurate localization
under limited spectral resources.

\section{Conclusion} 

In this paper, we experimentally investigated near-field target localization by reusing communication-centric OFDM pilot signals in a distributed multistatic ISAC system. The results demonstrate that accurate localization can be achieved with a limited number of distributed transmitters and receivers and moderate occupied bandwidth, without relying on large antenna arrays or dedicated radar waveforms. In particular, the fusion of bistatic excess-delay and receiver-pair TDOA information highlights the value of exploiting complementary geometric measurements across spatially distributed sensing links. The experimental findings also emphasize the importance of available bandwidth and practical calibration for reliable localization. Overall, this work demonstrates the feasibility of integrating sensing functionality into existing communication waveforms and provides insights into resource-efficient multistatic localization for future ISAC systems.

\bibliographystyle{IEEEtran}
\bibliography{IEEEabrv,ref}
\end{document}